\documentclass[11pt]{article}

\usepackage{amssymb}
\usepackage{pstricks}
\usepackage{multicol}
\usepackage{graphicx,graphics}
\usepackage{amsfonts}
\usepackage{color}
\newcommand{\rem}[1]{}
\DeclareMathAlphabet{\mathbi}{OML}{cmm}{b}{it} 
\newcommand{\non}{\nonumber}
\newtheorem{theorem}{Theorem}
\newtheorem{lemma}{Lemma}
\newtheorem{proposition}{Proposition}

\newcommand{\bx}{\mathbi{x}}

\newcommand{\bel}{\begin{equation}\label}
\newcommand{\ee}{\end{equation}}
\newcommand{\beq}{\begin{eqnarray}\label} 
\newcommand{\eeq}{\end{eqnarray}} 
\newcommand{\bc}{\begin{center}} 
\newcommand{\ec}{\end{center}} 
\newcommand{\ben}{\begin{enumerate}}
\newcommand{\een}{\end{enumerate}}
\newcommand{\bit}{\begin{itemize}}
\newcommand{\eit}{\end{itemize}}
\newcommand{\I}{\int_{\mathcal{V}}}

\newcommand{\bdf}{\mathbi{f}}

\newcommand{\bu}{\mathbi{u}}

\newcommand{\bw}{\mathbi{w}}

\newcommand{\bom}{\mbox{\boldmath$\omega$}}

\newcommand\shalf{\ensuremath{{\scriptstyle\frac{1}{2}}}}
\newcommand\squart{\ensuremath{{\scriptstyle\frac{1}{4}}}}

\newcommand\sevenfour{\ensuremath{{\scriptstyle\frac{7}{4}}}}

\newcommand{\Rey}{Re}
\newcommand{\Gr}{Gr}
\newcommand{\mG}{\mathcal{G}}
\newcommand{\mB}{\mathcal{B}}
\newcommand{\Y}{{\color{blue}\rule[0.0cm]{0.36cm}{.5mm}}}
\makeatletter
\@addtoreset{equation}{section}

\makeatother

\begin{document}
\sf
\bc
\par\vspace{1mm}
\textbf{\large Regularity and singularity in solutions of the\\
three-dimensional Navier-Stokes equations}
\ec
\bc
\textbf{\large J. D. Gibbon}
\par\vspace{2mm}
Department of Mathematics,
\par\vspace{2mm}
Imperial College London SW7 2AZ, UK
\par\vspace{2mm}
\textit{\small email: j.d.gibbon@ic.ac.uk}\\
%office phone: +44-207-594-8504~~home phone: +44-208-560-4353
\ec
\par\vspace{5mm}
\begin{abstract}
Higher moments of the vorticity field $\Omega_{m}(t)$ in the form of  $L^{2m}$-norms 
($1 \leq m < \infty$) are used to explore the regularity problem for solutions 
of the three-dimensional incompressible Navier-Stokes equations on the domain 
$[0,\,L]^{\,3}_{\,per}$.  It is found that the set of quantities
$$
D_{m}(t) = \Omega_{m}^{\alpha_{m}}\,,\qquad\qquad\alpha_{m} = \frac{2m}{4m-3}\,,
$$ 
provide a natural scaling in the problem resulting in a bounded set of time averages 
$\left<D_{m}\right>_{T}$ on a finite interval of time $[0,\,T]$. The behaviour of 
$D_{m+1}/D_{m}$ is studied on what are called `good' and `bad' intervals of $[0,\,T]$ 
which are interspersed with junction points (neutral) $\tau_{i}$. For large but finite 
values of $m$ with large initial data \big($\Omega_{m}(0) \leq \varpi_{0}O(\Gr^{4})$\big), 
it is found that there is an upper bound 
$$
\Omega_{m} \leq c_{av}^{2}\varpi_{0}\Gr^{4}\,,\qquad\varpi_{0} = \nu L^{-2}\,,
$$ 
which is punctured by infinitesimal gaps or windows in the vertical walls between the 
good/bad intervals through which solutions may escape. While this result is consistent 
with that of Leray \cite{Leray} and Scheffer \cite{Scheff76}, this estimate for $\Omega_{m}$ 
corresponds to a length scale well below the validity of the Navier-Stokes equations. 
\end{abstract}

\par\vspace{12mm}
%\tableofcontents
\bc 
3rd/09/09 (nsreg3.tex)
\ec

\thispagestyle{empty}
\newpage

%%%%%%%%%%%%%%%%%%%%%%%%%%%%%%
\section{\label{intro}\large\textsf{Introduction}}

The challenge that analysts have faced in the last 75 years has been to prove the existence 
and uniqueness of the three-dimensional Navier-Stokes equations for arbitrarily long times 
\cite{Leray,Lady,Serrin,CF,Temam,FMRT}. Its inclusion in the AMS Millenium Clay Prize list 
\cite{ClayFeff} has widely advertised the the nature of the problem but the elusiveness of 
a rigorous proof\footnote{\sf Cao and Titi \cite{CTprim} and Kobelkov \cite{Kob06} have 
recently proved the regularity of the primitive equations of the atmosphere and oceans, 
even though these have been considered by many to be a problem harder than the Navier-Stokes 
equations. The methods used unfortunately do not appear to successfully transfer to the 
Navier-Stokes equations.} and the severe resolution difficulties encountered in CFD, even 
at modest Reynolds numbers, are puzzles that have grown as the years progress.
\par\smallskip
Nevertheless, there is a long-standing belief in many scientific quarters, on the level of 
a folk-theorem, that the three-dimensional Navier-Stokes equations `must' be regular. 
Mathematicians are more cautious and still take seriously the possibility that singularities 
may occur, at least in principle. Leray \cite{Leray} and Scheffer \cite{Scheff76} proved that 
the (potentially) singular set in time has zero half-dimensional Hausdorff measure \cite{RS07}. 
The Leray-Scheffer result motivated Caffarelli, Kohn and Nirenberg \cite{CKN} to 
introduce the idea of suitable weak solutions to study the singular set in space-time which 
they concluded has zero one-dimensional Hausdorff measure. Thus, if space-time singularities 
exist then they must be relatively rare events. These ideas have spawned a growing literature 
on the subject where more efficient routes to the construction of suitable weak solutions are 
in evidence \cite{FLin,TX99,LadySeregin,ChoeLewis,Seregin03,He04,Seregin05,GG06,Seregin06}.
\par\smallskip
It is worth remarking that the wider issue regarding the formation of singularities has been 
obscured by the very great difficulty that exists in distinguishing them from rough intermittent 
data. Intermittency is characterized by violent surges or bursts away from averages in the 
energy dissipation, resulting in the spiky data that is now recognized as a classic hallmark of 
turbulence \cite{BT49,Kuo71,MS91,DCB,Lathrop}. At least three options are possible: 
\bit\itemsep -1mm
\item[a)] Solutions are always smooth with only mild excursions away from space and time averages; 

\item[b)] Solutions are intermittent but, despite their apparent spikiness, remain smooth for 
arbitrarily long times when examined at very small scales;

\item[c)] Solutions are intermittent but spikes may be the manifestation of true singularities.
\eit
Options b) and c) are impossible to distinguish using known computational methods. The 
Leray-Scheffer result shows that potential singularities in time must be distributed as no more 
than points on the time axis, but it contains little other information. Both for analytical and 
computational reasons it would be desirable to understand the origin of these points and the 
structure of the solution near to them. The aim of this paper is to address this issue.

In the past generation physicists have used Kolmogorov's theory to examine intermittent events 
by studying anomalies in the scaling of velocity structure functions. This theory is based on 
a set of statistical axioms, not directly on the Navier-Stokes equations. Nevertheless, to make 
a comparison, the intermittent dynamics discussed above would lie deep in the dissipation range 
of the energy spectrum. Frisch's book \cite{Frischbk} and the recent review by Boffetta, Mazzino 
and Vulpiani \cite{BMV} contain readable accounts of these ideas. 

%%%%%%%%%%%%%% removed piece %%%%%%%%%%%%%%%%%%%%%
\rem{\par\smallskip
Since Leray's pioneering work in the 1930's, the extensive use of the Sobolev-Gagliardo-Nirenberg-type 
inequalities has fallen just short of establishing a full rigorous proof of regularity. The history 
of the subject shows (see \cite{CF,Temam}) that to jump the gap between what is assumed and what is 
known\footnote{\sf Constantin and Fefferman have looked at the direction of vorticity to establish 
geometric regularity criteria \cite{ConstFeff,CFM}.}, the divergence-free velocity vector field 
$\bu(\bx,\,t)$ must be assumed to be \textit{a priori} bounded in either the $L^{p}$ ($p\geq 3$) 
or $H_{1}$-norms. The first of these assumptions defies physical interpretation but the second 
assumption, namely that of the bounded-ness of the point-wise-in-time global enstrophy defined 
below by ($\mbox{div}\,\bu = 0$)
\bel{H1def}
H_{1}(t)=\I|\nabla\bu|^{2}\,dV = \I|\bom|^{2}\,dV\,,
\qquad\qquad\bom = \mbox{curl}\,\bu\,,
\ee
is critical to the interpretation of experimental data \cite{BT49,Kuo71,MS91,DCB,Lathrop}. 
$H_{1}$ is proportional to the energy dissipation rate which Leray showed has an 
\textit{a priori} bounded time-average \cite{Leray}. Thus $H_{1}(t)$ is the key to the 
problem\,: it is a well-known fact that there exists a very short interval of time 
$[0,\,t_{0})$ on which solutions exist and are unique (see \cite{CF,Temam,FMRT}), but 
no proof exists to date that demonstrates this for arbitrarily long times\footnote{\sf The 
$\alpha$-model of Foias, Holm and Titi \cite{FHT2} has the $H_{1}$-norm of solutions 
bounded for all time. Control over $H_{1}$ is the fundamental feature of the regularization 
inherent in this model.}. The upper bound on $H_{1}(t)$ blows up at some short time $t_0$ 
at which time control is lost. Strong intermittency in the energy dissipation, even for 
modest Reynolds numbers, makes its ultimate fate uncertain.} 
%

%%%%%%%%%%%%%%%%%%%%%%%%%%%%%%%%%%%%%%%%%%%%%%%%%%%%%%%%%%%%
\subsection{\label{strategy}\large\textsf{General strategy}}

The main idea of this paper is to use higher moments of the vorticity field $\bom$ instead 
of derivatives. Scaled by a system volume term $L^{-3}$, a set of moments with the 
dimension of a frequency are defined such that for $m \geq 1$
\bel{Omintro1}
\Omega_{m}(t)  = \left\{L^{-3}\I|\bom|^{2m}\,dV\right\}^{1/2m} + \varpi_{0}\,,
\ee 
where $\varpi_{0} = \nu L^{-2}$ is the basic frequency of the domain of side $L$. $\Omega_{1}$ 
is synonymous with the $H_{1}$-norm and sits within the sequence of inequalities
\bel{sequ1}
\varpi_{0} < \Omega_{1}(t) \leq \Omega_{2}(t) \leq \ldots \leq \Omega_{m}(t) 
\leq \Omega_{m+1}(t) \leq \ldots,
\ee
so control from above over $\Omega_{m}$ for any value of $m > 1$ implies control over the 
$H_{1}$-norm which, in turn, controls from above all derivatives of the velocity field 
\cite{Lady,Serrin,CF,Temam,FMRT}. 

A technical problem lies in how to differentiate the $\Omega_{m}(t)$ and manipulate them without 
the existence of strong solutions for arbitrarily large $t$. This difficulty can be circumvented 
my restricting estimates to a finite interval of time $[0, T]$ and then pursuing a contradiction 
proof in the following standard manner. Assume that there exists a \textit{maximal} interval of 
time $[0, T_{max})$ on which solutions exist and are unique; that is, strong solutions are assumed 
to exist in this interval. If $[0, T_{max})$ is indeed maximal then $\Omega_{1}(T_{max}) = \infty$. 
The ultimate aim of such a calculation would then be to show that $\limsup_{T\to T_{max}}\Omega_{m}$ 
is finite for any $m \geq 1$; if this turned out to be the case it would lead to a contradiction 
because $[0, T_{max})$ would not be maximal. Thus $T_{max}$ must either be zero or infinity\,: it 
cannot be zero because it is known that there exists a short interval $[0,\,t_{0})$ on which strong 
solutions exist, so $T_{max}=\infty$. 

The results in \S\ref{Omsect} have been estimated using this strategy. It turns out that there exists 
a natural scaling within the Navier-Stokes equations which makes the variable
\bel{Dmdefintro}
D_{m}(t) = \left(\varpi_{0}^{-1}\Omega_{m}\right)^{\alpha_{m}}\qquad\mbox{with}\qquad
\alpha_{m} = \frac{2m}{4m-3}\,,
\ee
the most natural to choose. Then Theorem \ref{avthm} shows that
\bel{Dmintro3}
\left<D_{m}\right>_{T} \leq c_{av}\Gr^{2} + O\big( T^{-1}\big)\,,
%\frac{L\nu^{-3}E_{0}}{T}\,.
\ee
with a uniform constant $c_{av}$. Two remarks are in order. Firstly it is not difficult to extract 
an estimate for a set of length scales from (\ref{Dmintro3}). Defining $\lambda_{m}^{-2\alpha_{m}} 
= \nu^{-\alpha_{m}}\left<\Omega_{m}^{\alpha_{m}}\right>_{T}$, this shows that
\bel{Dmintro4}
\left(L\lambda_{m}^{-1}\right)^{2\alpha_{m}} = \left<D_{m}\right>_{T}\,,
\ee
and therefore\footnote{\sf Doering and Foias \cite{DF02} have shown that for Navier-Stokes
solutions $\Gr \leq c\,\Rey^{2}$ which would be valid if solutions were assumed to exist for large enough
values of $T$. In this case the $\Gr^{2}$-term on the right hand side of (\ref{Dmintro3}) would be replaced
by $\Rey^{3}$ in which case the right hand side of (\ref{Dmintro5}) would be $\Rey^{3/2\alpha_{m}}$. Thus,
$L\lambda_{1}^{-1} \leq c^{1/4}\Rey^{3/4}$ which is the Kolmogorov estimate. For large $m$, this becomes 
significantly larger running to $L\lambda_{m}^{-1} \leq c\,\Rey^{3}$.} 
\bel{Dmintro5}
L\lambda_{m}^{-1} \leq \big(c_{av}\Gr^{2}\big)^{1/2\alpha_{m}} + O\big(T^{-1}\big)\,.
\ee
\rem{Secondly, the elegance of these results cannot hide the fact that they are not enough to prove full 
regularity. To demonstrate this, drop the negative term on the right hand side of (\ref{Dmintro1}) 
which allows a direct time integration
\bel{Dmintro6}
D_{m}(T) \leq D_{m}(0)\exp \int_{0}^{T}\!D_{n}^{2}\,dt\,.
\ee
%An estimate for $\left<D_{m}^{\delta}\right>_{T}$ with $\delta = 2$ is needed.  
Unfortunately an estimate for $\left<D_{1}^{2}\right>_{T}$ is unavailable so the problem remains open. 
The need for a factor of 2 within the average has been known for some decades \cite{CF,Temam,FMRT}.}

The exponent $\alpha_{m}$ within the definition of $D_{m}$ appears to be a natural scaling consistent 
with that of the Sobolev inequalities. This paper suggests that the breaking of this scaling through 
stretching between $D_{m+1}$ and $D_{m}$ may be required to make progress. This is gauged more 
specifically in Theorem \ref{interthmA} in \S\ref{Omsect} where it is shown that a finite interval 
$[0,\,T]$ of the time axis can be potentially broken down into three classes, denoted by \textit{good} 
and \textit{bad} intervals with set of junction points (or intervals) $\{\tau_{i}\}$ designated as 
\textit{neutral}. In \S\ref{traj}, it is found that the direction of the inequality is 
reversed on the good and bad intervals; that is
\bel{interintro1}
\frac{D_{m+1}}{D_{m}} \lessgtr c_{av}D_{m}^{-\mu_{m}}\Gr^{p(T)}
\qquad\left\{\begin{array}{l}
<\hbox{(good)}\\
= \hbox{(neutral)}\\
>\hbox{(bad)}
\end{array}\right.
\ee
In (\ref{interintro1}) $p(T)$ is a $T$-dependent exponent ($>2$) of the Grashof number $\Gr$ and 
$\mu_{m}$ is a parameter in the range $0 < \mu_{m}< 1$. The universal inequality $\Omega_{m} \leq 
\Omega_{m+1}$ ultimately shows that on good and neutral intervals
\bel{Ombd1}
D_{m} \leq \mG_{m}^{\alpha_{m}}\,,
\ee 
where $\mG_{m}$ is a function of $p(T),\,\Gr,\,\alpha_{m}$ and $\mu_{m}$. The main question lies in the 
nature of the transition from the good to the bad intervals through the neutral points $\tau_{i}$.  
On bad intervals the application of the reverse inequality in (\ref{interintro1}) to the differential
inequality for $D_{m}$ in Proposition \ref{Dmthm} results in regions smaller in amplitude than $\mG_{m}$ 
in which solution trajectories remain bounded by 
\bel{Ombd2}
D_{m} \leq \mB_{m}^{\alpha_{m}}\,.
\ee
The bad regions are not absorbing\,: solutions remain inside these regions if they enter inside, but 
they are not attracted into them if they lie outside. The key point is that for all \textit{finite} 
values of $m\geq 1$, $\mB_{m} < \mG_{m}$, thereby leaving vertical gaps or windows through which 
trajectories can potentially escape to infinity -- see Figures 1,\,2 and 3. \textbf{However, while 
the gap between $\mG_{m}$ and $\mB_{m}$ closes for large $m$, the limit $m = \infty$ is forbidden 
and so these windows can only be reduced to infinitesimally small holes which puncture a general 
upper bound.} This result is consistent with that of Leray \cite{Leray} and Scheffer \cite{Scheff76}. 
In terms of $\Omega_{m}$, this punctured bound turns out to be 
\bel{Ombd3}
\Omega_{m} \lesssim c_{av}^{2}\varpi_{0}\Gr^{4}
\ee
When converted into a length scale, this estimate shows that regular solutions may go as deep as near 
nuclear scales  ($10^{-2}$ angstroms) and therefore many orders of magnitude below the validity of the 
Navier-Stokes equations.  The conclusion is that unless other unknown controlling mechanisms are shown 
to exist, the Navier-Stokes equations may formally possess solutions that either become singular or, if 
they continue to exist, may be unresolvable numerically.

%%%%%%%%%%%%%%%%%%%%%%%%%%%%%
\subsection{\label{notation}\large\textsf{Notation and functional setting}}

The setting is the incompressible ($\rm{div}\,\bu = 0$), forced, three-dimensional Navier-Stokes 
equations for the velocity field $\bu(\bx,\,t)$
\bel{NS1}
\bu_{t} + \bu\cdot\nabla\bu = \nu\Delta\bu - \nabla p + \bdf(\bx)\,,
\ee
with the equation for the vorticity expressed as
\bel{ns1}
\bom_{t} + \bu\cdot\nabla\bom = \nu\Delta\bom + \bom\cdot\nabla\bu + \hbox{curl}\bdf\,.
\ee
The properties of the forcing \& other definitions are given in Table \ref{tab1}. The domain 
$\mathcal{V}= [0,\,L]^{3}$ is taken to be three dimensional and periodic.  The forcing function 
$\bdf(\bx)$ is $L^2$-bounded and the Grashof number $\Gr$ is proportional to $\|\bdf\|_{2}$\,: 
see the paper by Doering and Foias \cite{DF02} for a discussion of narrow-band forcing 
\cite{DF02}\,: for simplicity the forcing is taken at a single length-scale $\ell =L/2\pi$.  
%%%%%%%%%%%%%%%%%%%%%%%%%%%%%%%%%%%%%%%%%%%
%%%%%%%%%%%%%%%%%%%%%
\begin{table}[ht]
\bc
\begin{tabular}{|l|l|l|} 
\hline\noalign{\smallskip}
%\extrarowheight{9pt}
\sf Quantity & \sf Definition & \sf Remarks\\
\noalign{\smallskip}
\hline
\noalign{\smallskip}
\sf Box length & $L$ &\\
\sf Forcing length scale & $\ell$ & $\ell = L/2\pi$\\
%\sf Forcing aspect ratio & $a_{\ell} = L/\ell$&\\
\sf Average forcing & $f_{rms}^{2} = L^{-3}\|\bdf\|_{2}^{2}$ &\\
\sf  Narrow-band forcing & $\|\bdf\|_{2}^{2} \approx \ell^{2n}\|\nabla^{n}\bdf\|_{2}^{2}$& $n\geq 1$\\
%\sf Average velocity & $U^{2}=L^{-3}\left<\|\bu\|_{2}^{2}\right>$ &\\
\sf Grashof No & $\Gr = \ell^{3}f_{rms}\nu^{-2}$ & \\
%Reynolds No  & $\Rey = U\ell\,\nu^{-1}$ & see \cite{DF02}\\
\sf Box frequency & $\varpi_{0} = \nu L^{-2}$ & \\
\sf Characteristic velocity & $u_{0} = L\varpi_{0}$ & \\
\sf $E$-definition & $E(t) = \I |\bu|^{2}\,dV$ & {\sf Energy}\\
%\sf Characteristic time & $\tau = \varpi_{0}^{-1}\Gr^{-1/2}$ & see \cite{DG02}\\
%\sf Characteristic velocity & $u_{0} = L\tau^{-1}$ & \\
%\sf Forcing frequency & $\bw_{f}(\bx) = u_{0}^{-1}\bdf(\bx)$ & \\
%\sf Forcing velocity & $\bu_{f}(\bx)  = \tau\bdf(\bx)$ & \\
%\sf Forcing length-scale & $\lambda_{m}^{-\frac{3}{2m(m+1)}} = 
%\|\bw_{f}\|_{2(m+1)}/\|\bw_{f}\|_{2m}$ & see (\ref{lam2})\\
%\sf $H_{n}$-definition & $H_{n} = \I |\nabla^{n}\bu|^{2}\,dV$ & \\
%\sf $F_{n}$-definition & $F_{n} = H_{n} + \tau^{2}\I |\nabla^{n}\bdf|^{2}\,dV$ & \\
\sf $\beta_{m}$-definition & $\beta_{m} = m(m+1)$ & \\
\sf $\alpha_{m}$-definition & $\alpha_{m} = \frac{2m}{4m-3}$ & \\
\sf $\rho_{m}$-definition & $\rho_{m} = 2m(4m+1)/3$ & \\
\noalign{\smallskip}\hline
\end{tabular}
\caption{\sf\label{tab1}\textit{\small Definitions of the main parameters. The 
forcing is taken at a single length-scale $\ell =L/2\pi$.}}  
\ec
\end{table}
\par\vspace{-4mm}\noindent
%%%%%%%%%%%%%%%%%%%%%%%%%%%%%%%%%%%%%%
Now define
\bel{Jmdef}
J_{m}(t) = \I |\bom|^{2m}dV\,,
\ee 
where the frequencies $\Omega_{m}$ are given by
\bel{Omdef2}
\Omega_{m}(t) = \big(L^{-3}J_{m}\big)^{1/2m} + \varpi_{0}\,.
\ee
The term $\varpi_{0}$ in (\ref{Jmdef}) provides a lower bound for $\Omega_{m}$. 
Indeed it is easy to prove that 
\bel{Omorder1}
\varpi_{0}\leq \Omega_{1}(t) \leq \Omega_{2}(t) \leq \ldots \leq \Omega_{m}(t)
\leq \Omega_{m+1}(t) \leq \ldots 
\ee
The symbol $\big<\cdot\big>_{T}$ denotes the time average up to time $T$
\bel{avdef}
\left<g(\cdot)\right>_{T} = \limsup_{g(0)}\frac{1}{T}\int_{0}^{T}g(\tau)\,d\tau\,.
\ee

%%%%%%%%%%%%%%%%%%%%%%%%%%%%%%%%%%%%%%%%

%%%%%%%%%%%%%%%%%%%
\section{\label{Omsect}\large\textsf{Some properties of the $\Omega_{m}(t)$}}

%%%%%%%%%%%%%%%%%%%
\subsection{\label{inequalav}\large\textsf{A differential inequality and a time average}}

This subsection firstly contains a result concerning the differential inequalities that 
govern the set of frequencies $\Omega_{m}(t)$. Secondly it contains a result that is 
an estimate for an upper bound on a set of time averages over the interval $[0,\,T]$. 
Finally it contains a result on the nature of exponential bounds on $[0,\,T]$. All of 
the proofs, which lie in Appendices A, B and C, are based on the contradiction strategy 
explained in \S\ref{strategy}. 
%\par\smallskip\noindent
Firstly we define 
\bel{Dmdef}
D_{m} = \big(\varpi_{0}^{-1}\Omega_{m}\big)^{\alpha_{m}}\qquad\qquad
\alpha_{m} = \frac{2m}{4m-3}\,.
\ee
\begin{proposition}\label{Dmthm}  
On $[O,T]$, for $1 \leq m < \infty$, $n = \shalf(m+1)$ and $\Gr \geq 1$, the $D_{m}$ satisfy 
\bel{Dmcor1} 
\left(\varpi_{0}\alpha_{m}\right)^{-1}\dot{D}_{m} \leq D_{m}\left\{
- \frac{1}{c_{1,m}}\left(\frac{D_{m+1}}{D_{m}}\right)^{\rho_{m}}D_{m}^{2} + 
c_{2,m}D_{n}^{2} + c_{3,m}\Gr\right\}\,,
\ee
where $\rho_{m} = 2m(4m+1)/3$. For the unforced case the last term on the right hand side 
of (\ref{Dmcor1}) is proportional to $c_{3,m}$. 
\end{proposition}
\par\vspace{2mm}\noindent
\textbf{Remark\,:} Note the strict inequality $m < \infty$\,: the Riesz transform used in the 
proof in Appendix \ref{proof1} requires the introduction of higher derivatives when $m=\infty$. 
\par\vspace{1mm}\noindent
\begin{theorem}\label{avthm}{\rm$\!$:}  For $1 \leq m \leq \infty$ and $\Gr \geq 1$
\bel{Omav}
\left<D_{m}\right>_{T} \leq c_{av}\left(\Gr^{2} + \frac{L^{-5}E_{0}}{\varpi_{0}^{3}T}\right)\,,
\ee
where $E_{0} = E(0)$ is the initial value of the energy. For the unforced case, the 
estimate is
\bel{unf1}
\left<D_{m}\right>_{T} \leq c\,\frac{L^{-5}E_{0}}{\varpi_{0}^{3}T}\,.
\ee
\end{theorem}
\par\vspace{2mm}\noindent
\textbf{Remark\,:} (\ref{Omav}) can also be expressed as 
\bel{Omavex1}
\left<D_{m}\right>_{T} \leq c_{av}\Gr^{p}\,,
\ee
where $C$ is a uniform constant. The $m$-independent exponent $p(T,\,E_{0},\Gr)$ written as 
\bel{pdef}
p(T,\,E_{0},\Gr) = 2 + \ln\left\{1 + \frac{L^{-5}E_{0}}{\varpi_{0}^{3}T}\Gr^{-2}\right\}(\ln\Gr)^{-1}\,.
\ee

%%%%%%%%%%%%%%%%%%%%%%%%%%%%%%%%%%%%%%%%%%%%%%%%%%%%%%%%%%%%%%%%
\section{\textsf{\label{traj}\large Trajectories on good, bad and neutral intervals}}

%%%%%%%%%%%%%%%%%%%%%%%%%%%%%%%%%%%%%%%%%%%%%%%%%%%%%%%%%%%%%%%%
\subsection{\textsf{\label{intervalbds}\large The ratio $D_{m+1}/D_{m}$}}

Given the result in Proposition \ref{Dmthm}, understanding the behaviour of the ratio 
$D_{m}/D_{m+1}$ is an important step.
\par\vspace{1mm}
\begin{theorem}\label{interthmA}
For the parameters $\mu_{m} = \mu_{m}(T,\,p,\,\Gr)$ with values in the 
range $0 < \mu_{m} < 1$, the ratio $D_{m}/D_{m+1}$ obeys the inequality
\bel{thm3a}
\left<\left[\frac{D_{m}}{D_{m+1}}\right]^{(1-\mu_{m})/\mu_{m}}
- \left[c_{av}^{-1}\Gr^{-p(T)}D_{m}^{\mu_{m}}\right]^{(1-\mu_{m})/\mu_{m}}
\right>_{T} \geq 0\,.
\ee
\end{theorem}
\textbf{Remark 1\,:} The proof lies in Appendix \ref{proof4} and is dependent 
on the result of Theorem \ref{avthm}.
\par\vspace{1mm}\noindent
\textbf{Remark 2\,:} Theorem \ref{interthmA} implies that while there must be intervals 
where the integrand is positive, there could also be intervals where it is negative. 
While it tells us nothing about the interval size or distribution it is clear that 
these are $T$-dependent. 
\par\vspace{1mm}
Formally the theorem leads to the conclusion that there exists at least one \textbf{good 
interval of time within $[0,\,T]$ on which\,:}
\bel{pf3d1A}
\frac{D_{m}}{D_{m+1}} >  \left[c_{av}\Gr^{p(T)}\right]^{-1}D_{m}^{\mu_{m}}\,,
\ee
while there potentially exist \textbf{bad intervals of time} on which 
\bel{pf3d1B}
\frac{D_{m}}{D_{m+1}} <  \left[c_{av}\Gr^{p(T)}\right]^{-1}D_{m}^{\mu_{m}}\,.
\ee
\textbf{Neutral points or intervals} represented\footnote{\sf There is no 
information on how the $\tau_{i}$ are distributed.} by the zeros of the integrand in 
(\ref{thm3a}) lying at 
\bel{taudef}
\tau_{i} = \tau_{i}(\mu_{m},\,p(T),\,\Gr)\,.
\ee 
In terms of $\Omega_{m+1}$ and $\Omega_{m}$ (\ref{pf3d1A}) and (\ref{pf3d1B}) become 
%\footnote{\sf The Heine-Borel Theorem says that an finite interval of $\mathbb{R}^{1}$ can be 
%covered by a countable union of open sets. The good and bad sets are open but the neutral points 
%are more difficult as they could be smeared out into finite intervals.}.
\beq{ibds4}
\frac{\Omega_{m+1}}{\Omega_{m}} &\lesseqgtr& \left(\mG_{m}
\varpi_{0}\Omega_{m}^{-1}\right)^{\gamma_{m}}
\qquad\left\{
\begin{array}{l}
\hbox{\textit{good}}\\
\hbox{\textit{neutral}}\\
\hbox{\textit{bad}}
\end{array}
\right.
\eeq
where $\mG_{m}$ and $\gamma_{m}$ are defined by
\bel{mGdef}
\mG_{m} = \left[c_{av}\Gr^{p(T)}\right]^{1/\big(\alpha_{m}\tilde{\mu}_{m}\big)}\,,
\ee
\bel{gammamdef2a}
\gamma_{m}\alpha_{m+1} = \alpha_{m}\tilde{\mu}_{m}\,,
\ee
\bel{tilmudef}
\tilde{\mu}_{m} = \mu_{m} - \frac{3}{m(4m+1)}\,.
\ee
The positivity of $\gamma_{m}$ requires that $\mu_{m}$ be bounded away from zero such that
\bel{mumdef}
\frac{3}{m(4m+1)} < \mu_{m} < 1\,.
\ee 
%%%%%%%%%%
%%%%%%%%%%
Because $\Omega_{m+1}\geq\Omega_{m}$, (\ref{ibds4}) shows that on \textbf{good and neutral intervals}
\bel{Ombd}
D_{m,\,good}\leq \mG_{m}^{\alpha_{m}}\qquad\qquad 1 \leq m \leq \infty\,.
\ee
\par\vspace{2mm}\noindent
Now we turn to the \textbf{bad intervals}\,: consider (\ref{pf3d1B}) in (\ref{Dmcor1}), in which case 
($\Omega_{n} \leq \Omega_{m}$)
\bel{Dmex2} 
\left(\varpi_{0}\alpha_{m}\right)^{-1}\dot{D}_{m} \leq D_{m}\left\{
- \frac{1}{c_{1,m}}\left(c_{av}\Gr^{p(T)}D_{m}^{-\mu_{m}}\right)^{\rho_{m}}D_{m}^{2}
+ c_{2,m}D_{m}^{2\alpha_{n}/\alpha_{m}} + c_{3,m}\Gr\right\}\,,
\ee
where $\rho_{m} = 2m(4m+1)/3$ but $m=\infty$ is forbidden. The range of validity of $\mu_{m}$ in (\ref{mumdef}) 
can be re-written as
%\bel{rhocon1}
$\rho_{m} > \mu_{m}\rho_{m} > 2$. 
%2/\rho < \mu_{m} < 1\,.\ee
Thus $\dot{D}_{m}\leq 0$ if, at the time of entry $\tau_{i}$ into a bad interval 
\bel{Dmen1}
\left(c_{av}\Gr^{p(T)}D_{m,bad}(\tau_{i})^{-\mu_{m}}\right)^{\rho_{m}}D_{m}^{2} 
\geq c_{1,m} c_{2,m}D_{m,bad}(\tau_{i})^{2\alpha_{n}/\alpha_{m}} + c_{3,m}\Gr\,.
\ee
Given that $\rho_{m}\mu_{m} > 2$ and $\alpha_{n} \geq \alpha_{m}$, the first term 
on the right hand side of (\ref{Dmen1}) is dominant. Using the lower bound $D_{m} 
\geq 1$ it is found that
\bel{Omen1}
D_{m,\,bad}(\tau_{i}) \leq \mB_{m}^{\alpha_{m}}\,,
\ee
where
\bel{mBdef}
\mB_{m} = \left\{\frac{1}{c_{1,m} c_{2,m}}
\left[c_{av}\Gr^{p(T)}\right]^{\rho_{m}}
- \frac{c_{3,m}}{c_{1,m}c_{2,m}}\Gr\right\}^{1/a_{m}}\,,
\ee
\bel{ambmdef}
a_{m} = 2(\alpha_{n} - \alpha_{m}) + \alpha_{m}\rho_{m}\mu_{m}\qquad\qquad
b_{m} = \alpha_{m}\rho_{m}\mu_{m}\,.
\ee

%%%%%%%%%%%%%%%%%%%%%%%%%%%%%%%%%%%%%%%%%%%%%%%%%%%%%%%%%%%%%%%%
\subsection{\textsf{\label{G1}\large How large are $\mG_{1}^2$ and $\mB_{1}^2$?}}

\par\vspace{-15mm}
%%%%%%%%%%%%%% Figure 1 %%%%%%%%%%%%%%%%%%%

\bc
\begin{minipage}[htb]{9cm}
\setlength{\unitlength}{.7cm}
\begin{picture}(11,11)(0,0)
%
%
%%%%% vertical grey lines %%%%%%%
%{\color{lightgray}\put(3.1,6){\rule[0.0cm]{.5mm}{1.5cm}}}
%{\color{gray}\put(3.5,6){\rule[0.0cm]{.5mm}{1.5cm}}}
%
%
%%%%%%%%% axes & coordinates %%%%%%%%%%%
\put(0,0){\vector(0,1){8}}
\put(0,0){\vector(1,0){10}}
\put(-1,7.5){\makebox(0,0)[b]{$D_{1}(t)$}}
\put(10.3,-.2){\makebox(0,0)[b]{$t$}}

%%%%%%%%%%% T and T^{*} %%%%%%%%%%%%%
%\put(9.3,-.5){\makebox(0,0)[b]{\small $T$}}
%\put(9.3,0){\vector(0,1){.2}}
%
%
%%%%%%%%%%%%%%%%%% labels %%%%%%%%%%%%%%%%%%%%
\put(-1,6){\makebox(0,0)[b]{$\Gr^{5}$}}
\put(-1,2.5){\makebox(0,0)[b]{$\Gr^{2}$}}
\put(0,3){\line(1,0){.1}}
%\put(8.5,1.6){\makebox(0,0)[b]{\small time~average}}
%\put(-1.2,2){\makebox(0,0)[b]{$\Gr^{2\zeta_{0}+1/2}$}}
%\put(2.7,-.5){\makebox(0,0)[b]{$t_{*}$}}
\put(1,5.2){\makebox(0,0)[b]{\small good}}
\put(4,2){\makebox(0,0)[b]{\small bad}}
\put(6,5.2){\makebox(0,0)[b]{\small good}}
\put(8,2){\makebox(0,0)[b]{\small bad}}
%
%%%%%%%%% vert lines %%%%%%%%%
%\put(3,0){\line(0,1){8}}
%\put(3.5,0){\line(0,1){8}}
%\put(7,0){\line(0,1){8}}
%\put(7.5,0){\line(0,1){8}}
%
%
%%%%%%%%% horiz lines & dots %%%%%%%%
\put(2.8,2.5){\line(1,0){2.3}}
\put(7,2.5){\line(1,0){2}}
\multiput(0,3)(.1,0){92}{.}
%
%
%%%% thick black horiz stripes %%%%%%%
\multiput(0,6)(.1,0){24}{\Y}
\multiput(5,6)(.1,0){16}{\Y}
\put(3,-.7){\makebox(0,0)[b]{\small $\tau_{1}$}}
\put(5.1,-.7){\makebox(0,0)[b]{\small $\tau_{2}$}}
\put(7.2,-.7){\makebox(0,0)[b]{\small $\tau_{3}$}}
%
%%%%% black horiz stripes %%%%%%%
%\put(3,6){\Z}
%\put(7,6){\Z}
%%%%%%%%%%%%%%%%%%%%%%%%%%
%
%
%%%%%%%%%%%%% bezier curves %%%%%%%%%%%%%%
\qbezier[400](0,0.5)(4,11)(7,1.5)
\qbezier[500](0,1.5)(2.5,-2)(3.01,7.70)
\qbezier[500](0,5)(6.8,-5)(7.1,7.8)
\end{picture}
\end{minipage}
\ec
\par\vspace{3mm}\noindent
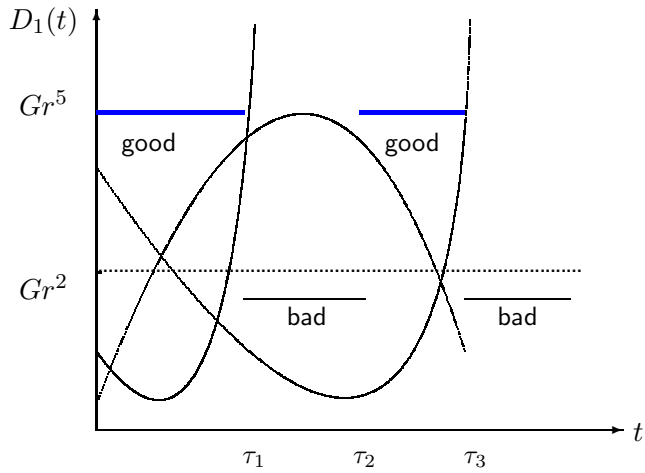
{\textbf{Figure 1\,:} \textit{\small From a variety of initial conditions for $m = 1$ 
the cartoon above shows how solutions may potentially escape at or near neutral points 
$t=\tau_{1}$ or a later value $t=\tau_{3}$, or even return at $t = \tau_{2}$. 
However, all must satisfy the bound on the time-average.}\label{mfig1}}
\par\vspace{5mm}\noindent
For $m=1$ we have $b_{1}/a_{1} = 1$ and $\rho_{1} = 10/3$\,; the difference in the sizes of $\mG_{1}$ 
and $\mB_{1}$ lies in the upper bounds on $\mu_{1}$ and on $\tilde{\mu}_{1}$. The latter has 
been defined in (\ref{tilmudef})
\bel{ubmu1}
\mu_{1} < 1\,,\qquad\qquad \tilde{\mu}_{1} < 1-3/5 = 2/5\,.
\ee
From (\ref{mGdef}) and (\ref{Ombd}) we have
\bel{ubmu2}
D_{1,\,good} \leq \left(c_{av}\Gr^{p}\right)^{1/\tilde{\mu}_{1}}
\ee
which, on minimization of the right hand side, gives
\bel{ubmu3}
D_{1,\,good} \leq \left(c_{av}\Gr^{2}\right)^{5/2} = c_{av}^{5/2}\Gr^{5}\,.
\ee
The equivalent estimate for $D_{1,\,bad}$ is 
\bel{ubmu4}
D_{1,\,bad} \leq \frac{c_{av}}{(c_{1,1}c_{2,1})^{3/10}}\Gr^{2} - O(\Gr^{3/10})\,.
\ee
It is useful to re-work these estimates in terms of a point-wise inverse\footnote{\sf The 
context of this is the estimate for the inverse length $L\lambda_{m}^{-1} \leq c_{av}^{1/4}
\Gr^{1/2}$ of \S\ref{intro}.} length-scale $\eta_{1}^{-4} = \nu^{-3}\epsilon$ with a 
point-wise energy dissipation rate $\epsilon = \nu\Omega_{1}^{2} = \nu\varpi_{0}^{2}D_{1}$. 
The result, 
\bel{ubmu5}
L\eta_{1}^{-1} \leq c_{av}^{1/4}\Gr^{1/2}
\ee
is shown in Figure 3 where the constant on the bad estimate is slightly smaller.

%%%%%%%%%%% Fig 2 %%%%%%%%%%

\par\vspace{-2cm}
\bc
\begin{minipage}[htb]{9cm}
\setlength{\unitlength}{.75cm}
\begin{picture}(11,11)(0,0)
%
%
%%%%% vertical grey lines %%%%%%%
%{\color{lightgray}\put(3.1,6){\rule[0.0cm]{.5mm}{1.5cm}}}
%{\color{gray}\put(3.5,6){\rule[0.0cm]{.5mm}{1.5cm}}}
%
%
%%%%%%%%% axes & coordinates %%%%%%%%%%%
\put(0,0){\vector(0,1){8}}
\put(0,0){\vector(1,0){10}}
\put(-1,7.5){\makebox(0,0)[b]{\small $L\eta_{1}^{-1}$}}
\put(10.2,0){\makebox(0,0)[b]{$t$}}

\put(1,5.2){\makebox(0,0)[b]{\small good}}
\put(4,1.5){\makebox(0,0)[b]{\small bad}}
\put(6,5.2){\makebox(0,0)[b]{\small good}}
\put(8,1.5){\makebox(0,0)[b]{\small bad}}
%
%%%%%%%%%%%%%%%%%% labels %%%%%%%%%%%%%%%%%%%%
\put(-1,6){\makebox(0,0)[b]{\small$\Gr^{5/4}$}}
\put(-1,2.6){\makebox(0,0)[b]{\small$\Gr^{1/2}$}}
%\put(-1,2.1){\makebox(0,0)[b]{\small$\Gr^{1/2}$}}
%\put(-1,2.8){\makebox(0,0)[b]{\small$\Gr^{3/8}$}}
\put(0,4.3){\line(1,0){.1}}
\put(5,2.5){\makebox(0,0)[b]{\small (time~average)}}
%\put(-1.2,3){\makebox(0,0)[b]{$\Gr^{2\zeta_{0}+1/2}$}}
%\put(2.7,-.5){\makebox(0,0)[b]{$t_{*}$}}
%
\put(3,-.7){\makebox(0,0)[b]{\small $\tau_{1}$}}
\put(5.1,-.7){\makebox(0,0)[b]{\small $\tau_{2}$}}
\put(7.2,-.7){\makebox(0,0)[b]{\small $\tau_{3}$}}
%
%
%%%%%%%%% vert lines %%%%%%%%%
%\put(3,0){\line(0,1){8}}
%\put(3.5,0){\line(0,1){8}}
%\put(7,0){\line(0,1){8}}
%\put(7.5,0){\line(0,1){8}}
%
%
%%%% thick black horiz stripes %%%%%%%
\multiput(0,6)(.1,0){24}{\Y}
\multiput(5,6)(.1,0){16}{\Y}
%
%
%%%%%%%%% horiz lines & dots %%%%%%%%
\put(2.85,2){\line(1,0){2.15}}
\put(7,2){\line(1,0){2.3}}
% horiz dotted line
\multiput(0,3)(.1,0){92}{.}
%
%
%%%%%%%%%%%%% bezier curves %%%%%%%%%%%%%%
%\qbezier[400](0,5)(3,1)(6,2)
\end{picture}
\end{minipage}
%$$
\par\vspace{7mm}\noindent
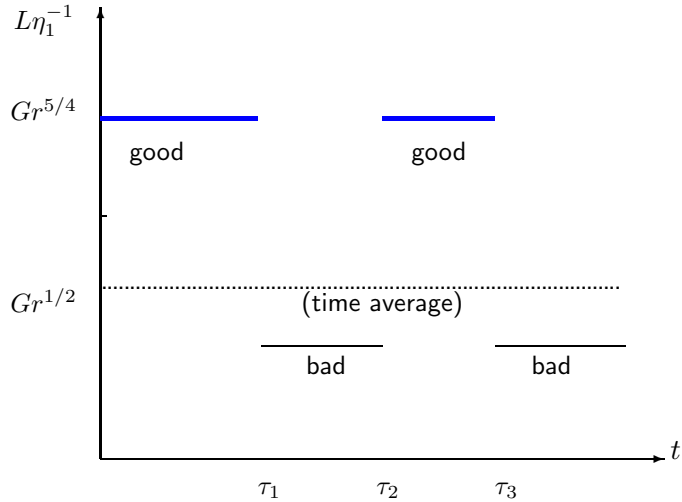
{\textbf{Figure 2\,:} \textit{\small Bounds on $L\eta_{1}^{-1}$\,: notice the large 
size of the gaps between the good and bad intervals. Based on the constants, 
the upper bound on the time average is larger than that on the bad intervals.}}
\ec
%

%%%%%%%%%%%%%%%%%%%%%%%%%%%%%%%%%%%%%%%%%%%%%%%%%%%%%%%%%%%%%%%%
\subsection{\textsf{\label{Gm}\large How large are $\mG_{m}^{\alpha_{m}}$ 
and $\mB_{m}^{\alpha_{m}}$ for large $m$?}}

%\par\vspace{2mm}\noindent

%%%%%%%%%%% Fig 3 %%%%%%%%%%

\par\vspace{-1.5cm}
\bc
\begin{minipage}[htb]{9cm}
\setlength{\unitlength}{.7cm}
\begin{picture}(11,11)(0,0)
%
%
%%%%%%%%% axes & coordinates %%%%%%%%%%%
\put(0,0){\vector(0,1){8}}
\put(0,0){\vector(1,0){10}}
\put(-1,7.5){\makebox(0,0)[b]{$D_{m}(t)$}}
\put(10.3,-.2){\makebox(0,0)[b]{$t$}}

%%%%%%%%%%% T and T^{*} %%%%%%%%%%%%%
%\put(9.3,-.5){\makebox(0,0)[b]{\small$T$}}
%\put(9.3,0){\vector(0,1){.2}}
%
\put(3,-.7){\makebox(0,0)[b]{\small $\tau_{1}$}}
\put(5.2,-.7){\makebox(0,0)[b]{\small $\tau_{2}$}}
\put(7.2,-.7){\makebox(0,0)[b]{\small $\tau_{3}$}}
%
%%%%%%%%%%%%%%%%%% labels %%%%%%%%%%%%%%%%%%%%
\put(-1,4){\makebox(0,0)[b]{$\Gr^{2}$}}
%\put(-1,3.4){\makebox(0,0)[b]{$\mB_{m}^{\alpha_{m}}$}}
%%
\put(4,2.5){\makebox(0,0)[b]{\small small~gaps through which}}
\put(4,2){\makebox(0,0)[b]{\small trajectories may pass}}
\put(2.6,3){\vector(0,1){.7}}
\put(6,3){\vector(-1,1){.7}}
\put(6,3){\vector(1,1){.7}}
%\put(2,3.7){\vector(1,-1){.7}}
%\put(1.5,2){\makebox(0,0)[b]{\small }}
%
%%%%%%%%%%%%% bezier curves %%%%%%%%%%%%%%
%\qbezier[400](2,4)(3.5,3)(4,3)
%
%\put(0,4.3){\line(1,0){.1}}
%\put(8,4.5){\makebox(0,0)[b]{\small time~average}}
%
%
%%%%%%%%% vert lines %%%%%%%%%
%\put(3,0){\line(0,1){8}}
%\put(3.5,0){\line(0,1){8}}
%\put(7,0){\line(0,1){8}}
%\put(7.5,0){\line(0,1){8}}
%
%
%%%% thick black horiz stripes %%%%%%%
\multiput(0,4.1)(.1,0){24}{\Y}
\multiput(5,4.1)(.1,0){16}{\Y}
%
%
%%%%%%%%% horiz lines & dots %%%%%%%%
\put(2.85,3.8){\line(1,0){2.15}} % 4.3
\put(7,3.8){\line(1,0){2.3}}
\multiput(0,3.9)(.1,0){92}{.}
\end{picture}
\end{minipage}
\ec
%$$
\par\vspace{4mm}\noindent
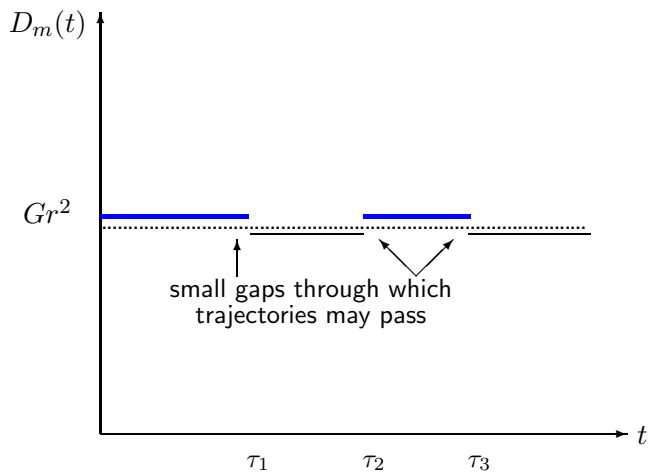
{\textbf{Figure 3\,:} \textit{\small For large $m$, the gap between $\mG_{m}^{\alpha_{m}}$ 
and $\mB_{m}^{\alpha_{m}}$ is infinitesimally small but the limit $m = \infty$ is forbidden. 
The upper bound on the time-average 
is the horizontal line of dots. At $\tau_{1}$ and $\tau_{3}$ a solution~must~enter the 
corresponding bad interval within the upper bound to remain inside.}\label{mfig3}}
\par\vspace{3mm}\noindent
From the definitions of (\ref{mGdef}) and (\ref{mBdef}) and the fact that $\tilde{\mu}_{m} 
< \mu_{m}$, it is clear that $\mG_{m}^{\alpha_{m}} -\mB_{m}^{\alpha_{m}} > 0$, keeping in 
mind that the limit $m = \infty$ is forbidden.  Firstly the $c_{i,m}$ are polynomial in 
$m$ and $\rho_{m} \sim O(m^{2})$ for large $m$. Therefore
\bel{Omen2}
\Big(c_{1,m}c_{2,m}\Big)^{-1/\rho_{m}} \nearrow 1\,,\qquad\qquad
b_{m}/a_{m} \nearrow 1\,,\qquad\mbox{and}\qquad
\mu_{m}^{-1}\nearrow \tilde{\mu}_{m}^{-1}\,.
\ee
Hence, for large $m$ 
\bel{BGlim}
\mG_{m}^{\alpha_{m}} - \mB_{m}^{\alpha_{m}} \searrow 0\,.
\ee
Specifically for $D_{m}$, for very large $m$, the upper bounds on $\mu_{m}$ and 
$\tilde{\mu}_{m}$ can now be taken arbitrarily close to unity provided that
$\mu_{m} < 1$ and $\tilde{\mu}_{m} < 1$\,.  From (\ref{mGdef}), minimization of 
the right hand side gives
\bel{mlg2}
D_{m,\,good} \leq c_{av}\Gr^{2}~~(\searrow)\,.
\ee
%where the right hand side limits down to the average $c_{av}\Gr^{2}$.  
The equivalent estimate for $D_{1,bad}$ is 
\bel{mlg3}
D_{m,\,bad} \leq \frac{c_{av}}{\big(c_{m,1}c_{m,2}\big)^{1/\rho_{m}}}\Gr^{2}
\nearrow c_{av}\Gr^{2}\,.
\ee

%%%%%%%%%%%%%%%%%%%%%%%%%%%%%
\section{\label{achieved}\large\textsf{Conclusion\,: what are the length scales 
corresponding to the upper bounds?}}

The key feature of this paper is the closure of the gaps between the good/bad 
intervals as $m\to\infty$ but with the actual limit $m = \infty$ forbidden.  
The origin of this lies in Proposition \ref{Dmthm} in the use of the inequality 
($p = \shalf (m+1)$)
\beq{Rinequal1}
\|\nabla\bu\|_{p} &\leq& c_{p}\|\bom\|_{p}\qquad\qquad p\in (1,\,\infty)\,,
\eeq
whereas, when $m = \infty$ 
\beq{Rinequal2}
\|\nabla\bu\|_{\infty} &\leq& c\,\|\bom\|_{\infty}\left(1 + \ln H_{3}\right)\,.
\eeq
(\ref{Rinequal1}) has its origin in a double Riesz transform while (\ref{Rinequal2}) arises 
from the work of Beale, Kato and Majda \cite{BKM} on the three-dimensional Euler equations  
-- see also Kato and Ponce \cite{KP86}.  The $\ln H_{3}$ term in (\ref{Rinequal2}) prevents 
the closure of the set of inequalities for $D_{m}$.  While the $m = \infty$ limit is valid 
for good intervals, it is not valid for the bad because of the necessary use of Proposition 
\ref{Dmthm}.  Thus it is not possible to completely close the gaps between the two sets of 
intervals, although they can become arbitrarily small. This allows for the possibility of 
the escape of trajectories. The $m$-dependence of the $\tau_{i}$ means that the junction 
points can, in principle, lie at different places on the time-axis as $m$ varies. If the 
gaps fall randomly with respect to $m$ then a trajectory would have to thread its way 
through these to escape to infinity. However, an unknown but subtle alignment of the 
gaps cannot entirely be ruled out.
\par
The closeness of the upper-bounds on both the time average and on point-wise values of 
$D_{m}$ ($m\gg 1$) away from the gaps, poses the question whether there exists dynamics 
that naturally lie either close to these bounds or even fulfill them. The point-wise 
energy dissipation rate per unit volume is 
\bel{varepdef1}
\varepsilon = \nu \Omega_{1}^{2} %= \nu^{3}L^{-4}D_{1}
\leq \nu^{3}L^{-4}D_{m}^{2/\alpha_{m}}\to \nu^{3}L^{-4}c_{av}^{4}\Gr^{8}\,.
\ee
Defining a \textit{local} Kolmogorov length as $\lambda_{k,loc} = \left(\varepsilon 
/\nu^{3}\right)^{1/4}$ we obtain
\bel{varepdef2}
L\lambda_{k,loc}^{-1} \leq c_{av}\Gr^{2}\,,
\ee
which is consistent with the estimate in (\ref{Dmintro4}) for large $m$. 
If the solution survives for large enough $T$ to make sense of a Reynolds number 
based on $U_{0}^{2} = L^{-3}\left<\|\bu\|_{2}^{2}\right>_{T}$, then the Doering-Foias 
result for Navier-Stokes solutions \cite{DF02}, $\Gr\leq c\,\Rey^{2}$, can be invoked 
to give an estimate for a \textit{local} Kolmorgorov 
scale\footnote{\sf The correspondence is that $\Gr^{2}$ is replaced by  $\Rey^{3}$.}
\bel{varepdef3}
L\lambda_{k,loc}^{-1} \leq c\,\Rey^{3}\,.
\ee
In the atmosphere, for instance, this length-scale would be of $O(10^{-12})$ metres -- 
about $10^{-2}$ angtroms -- which is about the scale of the nucleus (!) and is thus 
outside the validity of the Navier-Stokes equations. \textit{Because the bounds on 
the good and 
bad intervals are very close to the time average then solutions could, in principle 
spend long periods of time close to this bound and remain regular, yet such a scale 
is not only unreachable computationally but is outside the validity of the NS equations.
Thus, a singularity is not necessary to produce unresolvable solutions.}

\rem{
%%%%%%%%%%% Fig 4 %%%%%%%%%%

\par\vspace{-1.5cm}
\bc
\begin{minipage}[htb]{9cm}
\setlength{\unitlength}{.7cm}
\begin{picture}(11,11)(0,0)
%
%
%%%%% vertical grey lines %%%%%%%
%{\color{lightgray}\put(3.1,6){\rule[0.0cm]{.5mm}{1.5cm}}}
%{\color{gray}\put(3.5,6){\rule[0.0cm]{.5mm}{1.5cm}}}
%
%
%%%%%%%%% axes & coordinates %%%%%%%%%%%
\put(0,0){\vector(0,1){8}}
\put(0,0){\vector(1,0){10}}
\put(-1,7.5){\makebox(0,0)[b]{$D_{m}$}}
\put(10.3,-.2){\makebox(0,0)[b]{$t$}}

%%%%%%%%%%% T and T^{*} %%%%%%%%%%%%%
%\put(9.3,-.5){\makebox(0,0)[b]{\small$T$}}
%\put(9.3,0){\vector(0,1){.2}}
%
\put(3,-.7){\makebox(0,0)[b]{\small $\tau_{1}$}}
\put(5.2,-.7){\makebox(0,0)[b]{\small $\tau_{2}$}}
\put(7.2,-.7){\makebox(0,0)[b]{\small $\tau_{3}$}}
%
%%%%%%%%%%%%%%%%%% labels %%%%%%%%%%%%%%%%%%%%
\put(-1,4){\makebox(0,0)[b]{$\Gr^{2}$}}
%\put(-1,3.4){\makebox(0,0)[b]{$\mB_{m}^{\alpha_{m}}$}}
%\put(4.1,2.8){\makebox(0,0)[b]{\small small~gaps}}
%\put(2,3){\vector(1,1){.7}}
%\put(6,3){\vector(1,1){.7}}
%\put(6,3){\vector(-1,1){.7}}
%\put(0,4.3){\line(1,0){.1}}
%\put(8,4.5){\makebox(0,0)[b]{\small time~average}}
%
%
%%%%%%%%% vert lines %%%%%%%%%
%\put(3,0){\line(0,1){8}}
%\put(3.5,0){\line(0,1){8}}
%\put(7,0){\line(0,1){8}}
%\put(7.5,0){\line(0,1){8}}
%
%
%%%%%%% thick black horiz stripes %%%%%%%
\multiput(0,4.1)(.1,0){22}{\Y}
\multiput(5,4.1)(.1,0){15}{\Y}
%
%%%%%%%%% horiz lines & dots %%%%%%%%
\put(2.78,3.8){\line(1,0){2.15}} % 4.3
\put(7,3.8){\line(1,0){2.3}}
\multiput(0,3.9)(.1,0){92}{.}
%
%%%%%%%%%%%%% bezier curves %%%%%%%%%%%%%%
\qbezier[500](0,0.6)(4.7,8)(5.5,2) % 3rd curve
\qbezier[500](0,3.5)(3,0)(3.02,7.70)  %1st curve
\qbezier[500](0,3)(7.2,-3)(7.3,7.8) % 2nd curve
\qbezier[400](0,2)(4,4)(8,1)
\end{picture}
\end{minipage}
\ec
%$$
\par\vspace{4mm}\noindent
{\textbf{Figure 4\,:} For high values of $m$, examples of three trajectories that pass 
through the gaps and one that doesn't.}
}
\par\vspace{5mm}\noindent
\textbf{Acknowledgements:} I would like to express very warm thanks to Claude Bardos, 
Matania Benartzi, Toti Daskalopoulos, Darryl Holm, Roger Lewandowski, Gustavo Ponce, 
James Robinson and Edriss Titi for discussions on this topic.

%\newpage
%%%%%%%%%%%%%% Appendices %%%%%%%%%%%%%%%%%%%

\appendix

%%%%%%%%%%%%%%%%%%%%%% A1 %%%%%%%%%%%
\section{\label{proof1}\large\textsf{Proof of Proposition \ref{Dmthm}}}

Consider the time derivative of $J_{m}$ defined in (\ref{Jmdef})
\bel{Hm1}
\frac{1}{2m}\dot{J}_{m} = \I |\bom|^{2(m-1)}\bom\cdot
\left\{\nu\Delta\bom + \bom\cdot\nabla\bu + \hbox{curl}\bdf\right\}\,dV\,.
\ee 
Bounds on each of the three constituent parts of (\ref{Hm1}) are dealt with in turn, 
culminating in a differential inequality for $J_{m}$. In what follows, $c_{m}$ is 
a generic $m$-dependent constant.
\par\medskip\noindent
\textbf{1)~The Laplacian term\,:} Let $\phi = \omega^{2} = \bom\cdot\bom$. Then
\beq{s1a}
\I |\bom|^{2(m-1)}\bom\cdot\Delta\bom\,dV &=& 
\I \phi^{m-1}\bom\cdot\Delta\bom\,dV\nonumber\\
&=&\I \phi^{m-1}\{\Delta(\shalf\phi) - |\nabla\bom|^{2}\}\,dV\nonumber\\
&\leq& \I \phi^{m-1}\Delta(\shalf\phi)\,dV\,.
\eeq
Using the fact that $\Delta(\phi^{m}) = m\{(m-1)\phi^{m-2}|\nabla\phi|^{2} + 
\phi^{m-1}\Delta\phi\}$ we obtain
\beq{s1ex1}
\I |\bom|^{2(m-1)}\bom\cdot\Delta\bom\,dV 
&\leq& -\shalf(m-1)\I\phi^{m-2}|\nabla\phi|^{2}\,dV + 
\frac{1}{2m}\I\Delta(\phi^{m})\,dV\nonumber\\
&=& -\frac{2(m-1)}{m^2}\I|\nabla\phi^{\shalf m}|^{2}\,dV\nonumber\\
&=& - \frac{2(m-1)}{m^2}\I|\nabla(\omega^{m})|^{2}\,dV\,.
\eeq
Thus we have 
\beq{s1ex2}
\I |\bom|^{2(m-1)}\bom\cdot\Delta\bom\,dV \leq \left\{
\begin{array}{cl}
-\I|\nabla\bom|^{2}]\,dV & m=1\,,\\
-\frac{2}{\tilde{c}_{1,m}}\I|\nabla A_{m}|^{2}\,dV & m\geq 2\,.
\end{array}
\right.
\eeq
where
\bel{c1def}
A_{m}= \omega^{m}\qquad\qquad\tilde{c}_{1,m} = \frac{m^2}{m-1}\,,
\ee
where there is equality for $m=1$. 
The negativity of the right hand side of (\ref{s1ex2}) is important. Both 
$\|\nabla A_{m}\|_{2}$ and $\|A_{m}\|_{2}$ will appear later in the proof.  
\par\medskip\noindent
\textbf{2)~The nonlinear term in (\ref{Hm1})\,:} The second term in 
(\ref{Hm1}) is
\beq{s2a}
\left|\I |\bom|^{2(m-1)}\bom\cdot(\bom\cdot\nabla)\bu\,dV\right| 
&\leq& c_{m}\left(\I |\bom|^{2(m+1)}\,dV\right)^{\frac{m}{m+1}}
\left(\I |\nabla\bu|^{m+1}\,dV\right)^{\frac{1}{m+1}}\non\\
&\leq& c_{m}\left(\I |\bom|^{2(m+1)}\,dV\right)^{\frac{m}{m+1}}
\left(\I |\bom|^{m+1}\,dV\right)^{\frac{1}{m+1}}
\eeq
where the inequality $\|\nabla\bu\|_{p} \leq c_{p} \|\bom\|_{p}$ for $p \in (1,\,\infty)$ has been 
used\footnote{\sf I am grateful to G. Ponce for pointing this result out to me. Note that the 
$m=\infty$ case is forbidden because an extra $\log H_{3}$-term is needed \cite{BKM,KP86}. It is this 
forbidden limit that ultimately prevents the closure of the gaps in the figures in \S\ref{traj}, which 
allows trajectories to escape.}\,: this can be proved in the following way\,: write $\bu = \hbox{curl}
(-\Delta)^{-1}\bom$. Therefore $u_{i,j} = R_{j} R_{i}\,\omega_{i}$ where $R_{i}$ is a Riesz transform.  
\par\smallskip\noindent
Together with (\ref{s1a}) this makes (\ref{Hm1}) into 
\bel{Hm2}
\frac{1}{2m}\dot{J}_{m} \leq -\frac{\nu}{\tilde{c}_{1,m}}\I|\nabla(\omega^{m})|^{2}\,dV 
+ c_{m}J_{m+1}^{\frac{m}{m+1}}J_{\shalf(m+1)}^{\frac{1}{m+1}}
+ \I|\bom|^{2(m-1)}\bom\cdot\hbox{curl}\bdf\,dV\,.
\ee
\par\medskip\noindent
\textbf{3)~The forcing term in (\ref{Hm1})\,:}  Now we use the narrow-band property of the 
forcing (see the Table in \S\ref{notation}) to estimate the last term in (\ref{Hm2})
\beq{f2}
\I|\bom|^{2(m-1)}\bom\cdot\hbox{curl}\bdf\,dV 
&=& \I|\bom|^{2(m-1)}\bom\cdot\hbox{curl}\,\bdf\,dV \nonumber\\
&\leq& \left(\I |\omega|^{2m}dV\right)^{(2m-1)/2m}
\left(\I |\nabla\bdf|^{2m}\,dV\right)^{1/2m}\,.
\eeq
However, by going up to at least 3-derivatives in a Sobolev inequality it can easily be shown 
that $\|\nabla\bdf\|_{2m} \leq c\,\|\bdf\|_{2}L^{\frac{3-5m}{2m}}$, because of the 
narrow-band property. (\ref{f2}) becomes 
\beq{f3}
\left|\I|\bom|^{2(m-1)}\bom\cdot\hbox{curl}\bdf\,dV\right|
&\leq& c\,\left(L^{3}\Omega_{m}^{2m}\right)^{\frac{2m-1}{2m}}\|\bdf\|_{2}L^{\frac{3-5m}{2m}}\non\\
&\leq& c\,\Omega_{m}^{2m-1} f_{rms} L^{2}\non\\
&\leq& c\,\Omega_{m}^{2m-1}L^{3}\varpi_{0}^{2}\Gr
\eeq
\par\medskip\noindent
\textbf{4)~A differential inequality for $J_{m}$\,:} Recalling that $A_{m} = \omega^{m}$ 
\bel{s3a}
J_{m+1} = \I |\bom|^{2(m+1)}\,dV = \I |A_{m}|^{2(m+1)/m}\,dV
= \|A_{m}\|_{2(m+1)/m}^{2(m+1)/m}\,.
\ee
A Gagliardo-Nirenberg inequality yields
\bel{s3b}
\|A_{m}\|_{\frac{2(m+1)}{m}} \leq c_{m}\,\|\nabla A_{m}\|_{2}^{3/2(m+1)}
\|A_{m}\|_{2}^{(2m-1)/2(m+1)} + L^{-\frac{3}{2(m+1)}}\|A_{m}\|_{2}\,,
\ee
which means that 
\bel{s3c}
J_{m+1} \leq c_{m}\left\{\left(\I|\nabla(\omega^{m})|^{2}\,dV
\right)^{3/2m}J_{m}^{(2m-1)/2m} + L^{-3/m} J_{m}^{\frac{m+1}{m}}\right\}\,.
\ee
With $\beta_{m} = m(m+1)$, (\ref{s3c}) can be used to form $\Omega_{m+1}$ 
\beq{Jm1a}
\Omega_{m+1} = \left(L^{-3}J_{m+1}\right)^{1/2(m+1)} +\varpi_{0} 
&\leq& c_{m}\left(L^{-1}\I|\nabla(\omega^{m})|^{2}\,dV + L^{-3}J_{m} + \varpi_{0}^{2m}\right)^{3/4\beta_{m}}\non\\
&\times&\left[\left(L^{-3}J_{m}\right)^{1/2m} + \varpi_{0}\right]^{(2m-1)/2(m+1)}
\eeq
which converts to
\bel{Jm1b}
c_{m}\left(L^{-1}\I|\nabla(\omega^{m})|^{2}\,dV + L^{-3}J_{m} + \varpi_{0}^{2m}\right)
\geq \left(\frac{\Omega_{m+1}}{\Omega_{m}}\right)^{4\beta_{m}/3}\Omega_{m}^{2m}\,.
\ee
This motivates us to re-write (\ref{Hm2}) as
\beq{Jm2}
\frac{1}{2m}\big(L^{-3}\dot{J}_{m}\big) &\leq& -\frac{\varpi_{0}}{\tilde{c}_{1,m}}
\left(L^{-1}\I|\nabla(\omega^{m})|^{2}\,dV + L^{-3} J_{m} + \varpi_{0}^{2m}\right)\nonumber\\
&+& c_{2,m}\big(L^{-3}J_{m+1}\big)^{\frac{m}{m+1}}
\Big(L^{-3}J_{\shalf(m+1)}\Big)^{\frac{1}{m+1}}\non\\
&+& c_{3,m} \varpi_{0}L^{-3} J_{m} + c_{4,3}\varpi_{0}^{2m+1} + c_{5,m}\varpi_{0}^{2}\Omega_{m}^{2m-1}\Gr\,.
\eeq
Converting the $J_{m}$ into $\Omega_{m}$ and using $\Gr \geq 1$ 
\beq{Jm5}
\dot{\Omega}_{m} &\leq& \Omega_{m}\left\{-\frac{\varpi_{0}}{c_{4,m}}
\left(\frac{\Omega_{m+1}}{\Omega_{m}}\right)^{4m(m+1)/3} 
+ c_{5,m}\left(\frac{\Omega_{m+1}}{\Omega_{m}}\right)^{2m}
\Omega_{\shalf(m+1)} + c_{6,m}\varpi_{0}\Gr\right\}
\eeq
Using a H\"older inequality on the central term on the right hand side (\ref{Jm5}) 
finally becomes
\bel{new4}
\dot{\Omega}_{m} \leq \Omega_{m}\left\{-\frac{\varpi_{0}}{c_{1,m}}
\left(\frac{\Omega_{m+1}}{\Omega_{m}}\right)^{\frac{4\beta_{m}}{3}} 
+ c_{2,m}\varpi_{0}^{-\frac{3}{2m-1}}\Omega_{\shalf(m+1)}^{\frac{2(m+1)}{2m-1}} 
+ c_{3}\varpi_{0}\Gr\right\}\,.
\ee
With no forcing the final term in (\ref{new4}) is proportional 
to $\varpi_{0}^2$. Converting to the dimensionless quantity $D_{m} = \left(\varpi_{0}^{-1}
\Omega_{m}\right)^{\alpha_{m}}$ already defined in (\ref{Dmdef}) with $\alpha_{m} = 2m/(4m-3)$, 
finally gives 
\bel{Dminequal} 
\left(\varpi_{0}\alpha_{m}\right)^{-1}\dot{D}_{m} \leq D_{m}\left\{
- \frac{1}{c_{1,m}}\left(\frac{D_{m+1}}{D_{m}}\right)^{2m(4m+1)/3}D_{m}^{2} + c_{2,m}D_{n}^{2} 
+ c_{3,m}\Gr\right\}
\ee
with $n = \shalf (m+1)$. \hfil $\blacksquare$

%%%%%%%%%%%%%%%%%%% Proof of Thm 1 %%%%%%%%%%%%%%%
\section{\label{proof2}\large\textsf{Proof of Theorem \ref{avthm}}}

There exists a result of Foias, Guillop\'e and Temam \cite{FGT}, which uses higher derivatives. 
Define $H_{n}$ for $n \geq 1$
\bel{Hndef}
H_{n} = \I |\nabla^{n}\bu|^{2}\,dV\,,
\ee 
together with an integration of Leray's energy inequality 
\bel{energyinequal2}
\varpi_{0}^{-2}L^{-3}\left<H_{1}\right>_{T} =
\left<D_{1}\right>_{T} %= \varpi_{0}^{-2}\left<\Omega_{1}^{2}\right>_{T} c\,a_{\ell}^{4}
\leq \Gr^{2} + \frac{L\nu^{-3}E_{0}}{T}\,.
\ee
Then the result of Foias, Guillop\'e and Temam \cite{FGT} for $n\geq 3$ is 
\bel{app1}
\left< H_{n}^{\frac{1}{2n-1}}\right>_{T}\leq c_{n}\nu^{\frac{2}{2n-1}}
L^{-1}\left[\Gr^{2} + \frac{L\nu^{-3}}{T}E_{0}\right]\,,
\ee 
where $E_{0} = E(0) = H_{0}(0)$ is the initial energy. In the unforced case 
\bel{app2}
\left< H_{n}^{\frac{1}{2n-1}}\right>_{T}\leq c_{n}\nu^{-\frac{6n-5}{2n-1}}\frac{E_{0}}{T}\,.
\ee
A Sobolev inequality gives 
\bel{Omsob1}
\|\bom\|_{2m} \leq c\,\|\nabla^{2}\bom\|_{2}^{a}\|\bom\|_{2}^{1-a} 
\ee
where $a = 3(m-1)/4m$ for $m \geq 1$.  Moreover, the constant $c$ can be taken as finite for 
each finite $m$ because the $m=\infty$ case it is a bounded. Thus, taking $n = 3$ in 
(\ref{app1}), which fixes the constant $c_{n}$, we have 
\beq{Omsob2a}
\left<\|\bom\|_{2m}^{\frac{2m}{4m-3}}\right>_{T} &\leq& 
c\,\left<\left(H_{3}^{1/5}\right)^{\frac{15(m-1)}{4(4m-3)}}H_{1}^{\frac{m+3}{4(4m-3)}}
\right>_{T}\nonumber\\
&\leq& c\,\left<\left(H_{3}^{1/5}\right)\right>_{T}^{\frac{15(m-1)}{4(4m-3)}}
\left<H_{1}\right>_{T}^{\frac{m+3}{4(4m-3)}}\,.
\eeq
Using (\ref{energyinequal2}) and (\ref{app2}) this gives
\bel{Omsob2b}
\left<\|\bom\|_{2m}^{\frac{2m}{4m-3}}\right>_{T} 
\leq c_{av}\nu^{\frac{2m}{4m-3}}\left(L^{-1}\Gr^{2} + \frac{\nu^{-3}}{T}E_{0}\right)\,,
\ee
and thus the final result with an $m$-independent constant. In the unforced case 
\bel{Omsob3}
\left<\|\bom\|_{2m}^{\frac{2m}{4m-3}}\right>_{T} \leq c\,\varpi_{0}^{\alpha_{m}}
\left(\frac{L^{-5}E_{0}}{T\varpi_{0}^{3}}\right)L^{3\alpha_{m}/2m}\,.
\ee
There is also a way of reproducing the $\Gr^{2}$-estimate from 
Proposition \ref{Dmthm} but with worse constants. 
Based on $\Omega_{n}^{m+1} \leq \Omega_{m}^{m}\,\Omega_{1}$ for $n = \shalf(m+1)$, the 
relation in terms of the $D_{n}$ and $D_{m}$ is 
\bel{Dnrel1}
D_{n}^{2} \leq D_{m}^{\frac{4m-3}{2m-1}}D_{1}^{\frac{1}{2m-1}}\,.
\ee
Inequality (\ref{Dminequal}) is now divided by $D_{m}^{2-\delta}$ where 
$\delta \geq \frac{1}{(2m-1)}$. Noting that $D_{m} \geq 1$ the $D_{n}^{2}$-term is 
handled as follows 
\beq{nterm1}
\left<D_{n}^{2}D_{m}^{\delta - 2}\right>_{T} &\leq& 
\left<\left(D_{m}^{\delta}\right)^{\frac{(2m-1)\delta - 1}{(2m-1)\delta}}
\left(D_{1}^{\delta}\right)^{\frac{1}{(2m-1)\delta}}\right>_{T}\non\\
&\leq&  \left(\frac{(2m-1)\delta - 1}{(2m-1)\delta}\right)\left<D_{m}^{\delta}\right>_{T} + 
\frac{1}{(2m-1)\delta}\left<D_{1}^{\delta}\right>_{T}\,.
\eeq
It follows that
\beq{Dmres2} 
\left<\left(\frac{D_{m+1}}{D_{m}}\right)^{2m(4m+1)/3}D_{m}^{\delta}\right> 
&\leq& c_{4,m}\left<D_{m}^{\delta}\right> + c_{5,m}\left<D_{1}^{\delta}\right> 
+ c_{6,m}\Gr + O\big(T^{-1}\big)
\eeq
where the coefficients from the H\"older inequality have been absorbed into the constants. 
Define $\Delta_{m} = 2m(4m+1)/3$, and consider
\beq{Dmres3}
\left<D_{m+1}^{\delta}\right> &=& 
\left<\left[\left(\frac{D_{m+1}}{D_{m}}\right)^{\Delta_{m}}D_{m}^{\delta}\right]^{\delta/\Delta_{m}}
(D_{m}^{\delta})^{\frac{\Delta_{m}-\delta}{\Delta_{m}}}\right>\non\\
&\leq& \frac{\delta}{\Delta_{m}}\left<\left(\frac{D_{m+1}}{D_{m}}\right)^{\Delta_{m}}D_{m}^{\delta}\right>
+ \left(\frac{\Delta_{m}-\delta}{\Delta_{m}}\right)\left<D_{m}^{\delta}\right>
\eeq
where a H\"older inequality has been used at the last step. The end result is 
\bel{Dmres4}
\left<D_{m+1}^{\delta}\right> \leq c_{7,m}\left<D_{m}^{\delta}\right> 
+ c_{8,m}\left<D_{1}^{\delta}\right> 
+ c_{9,m}\Gr + O\big(T^{-1}\big)\,.
\ee
Because $n = \shalf(m+1)$, when $m=1$ then $n=1$. Moreover, only when $\delta = 1$ 
does an estimate exist for $\left<D_{1}\right>$ through (\ref{energyinequal2}), then 
(\ref{Dmres4}) is a generating inequality gives the $\Gr^{2}$-estimate but with worse 
constants. \hfill $\blacksquare$

\rem{
%%%%%%%%%%%%%%%%%%%%%%%%%%%%%%%%%%%%%%%%
\subsection{Standard proof}

Although\footnote{\sf Another route to the proof of Theorem \ref{avthm} is through the result 
of Foias, Guillop\'e and Temam \cite{FGT}. However, we have pursued the route taken in this 
paper to show that the proof is possible without the use of higher derivatives} it is strictly 
true that $\Omega_{\shalf(m+1)}\leq \Omega_{m}$ a sharper result is found by using Cauchy's 
inequality to obtain $\Omega_{\shalf(m+1)}^{m+1}\leq \Omega_{m}^{m}\Omega_{1}$. Therefore the 
nonlinear term in (\ref{Omcor1}) can be re-written as
\bel{Omhalf}
\Omega_{m}\Omega_{\shalf(m+1)}^{\frac{2m+2}{2m-1}} 
\leq \Omega_{m}^{1+\frac{2m}{2m-1}}\Omega_{1}^{\frac{2}{2m-1}}\,.
\ee
The first step is to divide (\ref{Omcor1}) by $\Omega_{m}^{1+ \alpha_{m}}$ where 
$\alpha_{m} = \frac{2m}{4m-3}$ and then average over $[0,\,T]$
\beq{newest1}
\varpi_{0}^{\alpha_{m}}\left<\frac{\Omega_{m+1}^{4\beta_{m}/3}}{\Omega_{m}^{4\beta_{m}/3}
\Omega_{m}^{\alpha_{m}}}\right>_{T} 
&\leq& c_{m}\varpi_{0}^{\alpha_{m}-\frac{2(m+1)}{2m-1}}
\left<\Omega_{m}^{\frac{2m}{2m-1} - \alpha_{m}}\Omega_{1}^{\frac{2}{2m-1}}\right>_{T} 
+ c_{6,m}\big(\varpi_{0}T\big)^{-1}\non\\ 
&+& c_{7,m}\big(\Gr + 1\big)\non\\
&\leq&
c_{m}\left[\varpi_{0}^{-\alpha_{m}}\left<\Omega_{m}^{\alpha_{m}}\right>_{T}\right]^{\frac{2m-2}{2m-1}}
\left[\varpi_{0}^{-2}\left<\Omega_{1}^{2}\right>_{T}\right]^{\frac{1}{2m-1}} 
+ c_{6,m}\big(\varpi_{0}T\big)^{-1}\non\\ 
&+& c_{7,m}\big(\Gr +1\big)\,.
\eeq
Given that $\alpha_{m} = \frac{2m}{4m-3}$, a key observation is that it satisfies the relation
\bel{formula1}
\frac{(4\beta_{m}+3\alpha_{m})\alpha_{m+1}}{4\beta_{m}-3\alpha_{m+1}} = \alpha_{m}\,.
\ee
Keeping this in mind, a manipulation of $\left<\Omega_{m+1}\right>_{T}$ gives
\beq{newest2}
\varpi_{0}^{-\alpha_{m+1}}\left<\Omega_{m+1}^{\alpha_{m+1}}\right>_{T} &=& 
\varpi_{0}^{-\alpha_{m+1}}\left<\left(\frac{\Omega_{m+1}^{4\beta_{m}/3}}{\Omega_{m}^{4\beta_{m}/3}
\Omega_{m}^{\alpha_{m}}}\right)^{3\alpha_{m+1}/4\beta_{m}}
\Omega_{m}^{\alpha_{m+1} + 3\alpha_{m}\alpha_{m+1}/4\beta_{m}}\right>_{T}\nonumber\\
&\leq& 
\left<\frac{\varpi_{0}^{\alpha_{m}}\Omega_{m+1}^{4\beta_{m}/3}}{\Omega_{m}^{4\beta_{m}/3}
\Omega_{m}^{\alpha_{m}}}\right>_{T}^{3\alpha_{m+1}/4\beta_{m}}
\left<\big(\varpi_{0}^{-1}\Omega_{m}\big)^{\alpha_{m}}\right>_{T}^{\frac{4\beta_{m}-3\alpha_{m+1}}{4\beta_{m}}}\non\\
&\leq& 
\left(\frac{3\alpha_{m+1}}{4\beta_{m}}\right)
\left<\frac{\varpi_{0}^{\alpha_{m}}\Omega_{m+1}^{4\beta_{m}/3}}{\Omega_{m}^{4\beta_{m}/3}
\Omega_{m}^{\alpha_{m}}}\right>_{T} + 
\left(\frac{4\beta_{m} - 3\alpha_{m+1}}{4\beta_{m}}\right)
\left<\big(\varpi_{0}^{-1}\Omega_{m}\big)^{\alpha_{m}}\right>_{T}\non
\eeq
which, together with (\ref{newest1}) and a H\"older inequality, becomes
\beq{newest6}
\varpi_{0}^{-\alpha_{m+1}}\left<\Omega_{m+1}^{\alpha_{m+1}}\right>_{T} &\leq& 
c_{8,m}\varpi_{0}^{-\alpha_{m}}\left<\Omega_{m}^{\alpha_{m}}\right>_{T}
+ c_{9,m}\varpi_{0}^{-2}\left<\Omega_{1}^{2}\right>_{T}\non\\
&+& c_{10,m}\big(\varpi_{0}T\big)^{-1} + c_{11,m}\big(\Gr+1\big)\,.
\eeq
An integration of Leray's energy inequality gives an estimate 
for $\left<L^{-3}\Omega_{1}^{2}\right>_{T}$ 
\bel{energyinequal2}
\varpi_{0}^{-2}\left<L^{3}\Omega_{1}^{2}\right>_{T} \leq %c\,a_{\ell}^{4}
\Gr^{2} + \frac{L\nu^{-3}E_{0}}{T}\,,
\ee
so (\ref{newest6}) can be used as a generating function to find the estimate for 
$\left<\Omega_{m}^{\alpha_{m}}\right>_{T}$ for all $m \geq 1$ as advertised.
\hfill $\blacksquare$

%%%%%%%%%%%%%%%%%%% Proof of Thm 2 %%%%%%%%%%%%%%%

\section{\label{proof3}\large\textsf{Proof of Theorem \ref{expthm}}}

Firstly, it should noted that
\beq{Ominequal1}
\I |\bom|^{m+1}dV &=& \I |\bom|^{\shalf(m+1)}|\bom|^{\shalf(m-1)}|\bom|\,dV\non\\
&\leq& \left(\I |\bom|^{2(m+1)}\,dV\right)^{1/4}
\left(\I |\bom|^{2(m-1)}\,dV\right)^{1/4}\left(\I |\bom|^{2}\,dV\right)^{1/2}
\eeq
and so 
\beq{Ominequal2}
\Omega_{\shalf(m+1)}^{2(m+1)} &\leq& \Omega_{m+1}^{m+1} \Omega_{m-1}^{m-1} \Omega_{1}^{2}\non\\
&=& \left(\frac{\Omega_{m+1}}{\Omega_{m}}\right)^{m+1}\Omega_{m}^{m+1}\Omega_{m-1}^{m-1} \Omega_{1}^{2}\,.
\eeq
Thus
\beq{mp1}
\left(\varpi_{0}^{-1}\Omega_{\shalf(m+1)}\right)^{\frac{2(m+1)}{2m-1}}
&\leq& \left(\frac{\Omega_{m+1}}{\Omega_{m}}\right)^{\frac{m+1}{2m-1}}
\left(\varpi_{0}^{-1}\Omega_{m}\right)^{\frac{m+1}{2m-1}}
\left(\varpi_{0}^{-1}\Omega_{m-1}\right)^{\frac{m-1}{2m-1}}
\left(\varpi_{0}^{-1}\Omega_{1}\right)^{\frac{2}{2m-1}}
\eeq
and so a H\"older inequality gives
\beq{mp2}
\left<\left[c_{m}\left(\varpi_{0}^{-1}\Omega_{\shalf(m+1)}\right)^{\frac{2(m+1)}{2m-1}}\right]^{1/2}\right>_{T}
&\leq& d_{1,m}\left<\left[\left(\frac{\Omega_{m+1}}{\Omega_{m}}\right)^{4\beta_{m}/3}\right]^{1/2}\right>_{T}
^{\frac{3(m+1)}{4\beta_{m}(2m-1)}}\non\\
&\times&d_{2,m}\left<\left(\varpi_{0}^{-1}\Omega_{m}\right)^{\alpha_{m}}\right>_{T}^{\frac{m+1}{2\alpha_{m}(2m-1)}}\non\\
&\times&\left<\left(\varpi_{0}^{-1}\Omega_{m-1}\right)^{\alpha_{m-1}}\right>_{T}^{\frac{m-1}{2\alpha_{m-1}(2m-1)}}\non\\
&\times&\left<\left(\varpi_{0}^{-1}\Omega_{1}\right)^{2}\right>_{T}^{\frac{1}{2(2m-1)}}
\eeq
provided 
\bel{mp3}
1 = \frac{3(m+1)}{4\beta_{m}(2m-1)} + \frac{m+1}{2\alpha_{m}(2m-1)} + \frac{m-1}{2\alpha_{m-1}(2m-1)}
+ \frac{1}{2(2m-1)}\,.
\ee
Using $\alpha_{m} =2m/(4m-3)$ it is easy to check that this identity is correct. Then (\ref{mp2}) can be 
re-written as
\beq{mp4}
\left<\left[c_{m}\left(\varpi_{0}^{-1}\Omega_{\shalf(m+1)}\right)^{\frac{2(m+1)}{2m-1}}\right]^{1/2}\right>_{T}
&\leq& d_{1,m}\frac{3(m+1)}{4\beta_{m}(2m-1)}
\left<\left[\left(\frac{\Omega_{m+1}}{\Omega_{m}}\right)^{4\beta_{m}/3}\right]^{1/2}\right>_{T}\non\\
&+&d_{2,m}\frac{m+1}{2\alpha_{m}(2m-1)}\left<\left(\varpi_{0}^{-1}\Omega_{m}\right)^{\alpha_{m}}\right>_{T}\non\\
&+&\frac{m-1}{2\alpha_{m-1}(2m-1)}\left<\left(\varpi_{0}^{-1}\Omega_{m-1}\right)^{\alpha_{m-1}}\right>_{T}\non\\
&+&\frac{1}{2(2m-1)}\left<\left(\varpi_{0}^{-1}\Omega_{1}\right)^{2}\right>_{T}
\eeq
Now choose $d_{1,m}$ such that $d_{1,m}\left(\frac{3(m+1)}{4\beta_{m}(2m-1)}\right) = 1$ and use the known
bounds on the last 3 terms on the right hand side to give
\beq{mp5}
\left<\left[\left(\frac{\Omega_{m+1}}{\Omega_{m}}\right)^{4\beta_{m}/3}\right]^{1/2}
- \left[c_{1,m}\left(\varpi_{0}^{-1}\Omega_{\shalf(m+1)}\right)^{\frac{2(m+1)}{2m-1}}\right]^{1/2}\right>_{T}
&+& c_{2,m}\Gr^{p} \geq 0
\eeq
where $p$ is defined in (\ref{pdef}). The square-roots inside the time average are unfortunate but are
a result of Navier-Stokes scaling properties.\hfill $\blacksquare$
}

%%%%%%%%%%%%%%%%%%% Proof of Thm 3 %%%%%%%%%%%%%%%
%
\section{\label{proof4}\large\textsf{Proof of Theorem \ref{interthmA}}}

With $0 < \mu_{m} < 1$ we write 
\beq{pf3a}
\left<D_{m}^{1-\mu_{m}}\right>_{T} &=& 
\left<\left(\frac{D_{m}}{D_{m+1}}\right)^{1-\mu_{m}}
D_{m+1}^{1-\mu_{m}}\right>\non\\
&\leq& 
\left<\left(\frac{D_{m}}{D_{m+1}}\right)^{\frac{1-\mu_{m}}{\mu_{m}}}\right>_{T}^{\mu_{m}}
\left<D_{m+1}\right>_{T}^{1-\mu_{m}}\,,
\eeq
which becomes
\beq{pf3b}
\left<\left(\frac{D_{m}}{D_{m+1}}\right)^{\frac{1-\mu_{m}}{\mu_{m}}}\right>_{T}
&\geq& \left(\frac{\left<D_{m}^{1-\mu_{m}}\right>_{T}}{\left<D_{m+1}\right>_{T}}
\right)^{\frac{1-\mu_{m}}{\mu_{m}}}\left<D_{m}^{1-\mu_{m}}\right>_{T}\,.
\eeq
The estimate for the time average of $\left<D_{m+1}\right>_{T}$ from (\ref{Omav}) and the 
lower bound $D_{m}\geq 1$ are now used to give
\beq{pf3c}
\left<\left[\frac{D_{m}}{D_{m+1}}\right]^{\frac{1-\mu_{m}}{\mu_{m}}} - 
\left[c_{av}^{-1}\Gr^{-p}D_{m}^{\mu_{m}}
\right]^{\frac{1-\mu_{m}}{\mu_{m}}}\right>_{T}\geq 0\,.
\eeq
This ends the proof of Theorem \ref{interthmA}. \hfill $\blacksquare$

%%%%%%%%%%%%%%%%%%%%%%%%%%%

\end{document}